# Discovery of a nonsymmorphic superconductor with spontaneous rotational symmetry breaking and nontrivial zero modes

Hui Guo[1,2,#], Zhixuan Li[2,#], Senhao Lv[1,2,#], Tianqi Gao[2,#], Zihao Huang[1,2], Kuanrong Hao[3], Lizhi Zhang[3], Ke Zhu[1,2], Siyu Li[1,2], Xianghe Han[1,2], Xiao Lin[2], Shengshan Qin[4], Wu Zhou[2], Haitao Yang[1,2], Hui Chen[1,2,*], and Hong-Jun Gao[1,2,*]

[1] Beijing National Center for Condensed Matter Physics and Institute of Physics, Chinese Academy of Sciences, Beijing 100190, China

[2] School of Physical Sciences, University of Chinese Academy of Sciences, Beijing 100190, China

[3] National Center for Nanoscience and Technology, Chinese Academy of Sciences, Beijing 100190, China

[4] School of Physics, Beijing Institute of Technology, Beijing 100081, China

[#]These authors contributed equally to this work.

*Correspondence to: hjgao@iphy.ac.cn, hchenn04@iphy.ac.cn

**Topological superconductivity has attracted great interest due to its fundamental significance for realizing Majorana quasiparticles and fault-tolerant quantum computation. Nonsymmorphic superconductors, with symmetry-protected nontrivial electronic structures, offer a promising route to exotic topological superconducting states, yet experimental realizations remain scarce. Here we identify nonsymmorphic compound $PtPb_4$ as a robust platform hosting superconductivity with spontaneous rotational symmetry breaking and nontrivial zero-energy modes. $PtPb_4$ crystallizes in a frustrated Shastry-Sutherland lattice and exhibits nontrivial band topology. By combining in-plane and out-of-plane resistivity measurements, pronounced twofold anisotropy is observed in both the superconducting state and the upper critical field, evidencing spontaneous rotational symmetry breaking. Scanning tunneling microscopy/spectroscopy further reveal twofold-symmetric magnetic vortices, providing direct real-space evidence for the symmetry-broken superconducting state. Notably, a robust zero-energy vortex bound state emerges and persists without spatial splitting over extended distances, consistent with the characteristics expected for Majorana bound state. These findings uncover an exotic superconducting state in $PtPb_4$ and establish a promising platform for exploring topological superconductivity and superconducting quantum devices.**

## Introduction

Topological superconductivity have emerged as a central frontier in condensed matter physics because of its fundamental significance for realizing non-Abelian quasiparticles for fault-tolerant topological quantum computation[1-4]. A defining hallmark of topological superconductivity is the emergence of Majorana zero modes as robust excitations localized at boundaries or vortex cores[5-14]. To date, experimental signatures of Majorana modes and topological superconductivity have been reported mainly in two classes of systems: engineered hybrid heterostructures based on proximity-induced superconductivity in topological materials[2,5-9,15], and a limited number of bulk superconductors with nontrivial band topology or exotic pairing states, including doped topological insulator $Bi_2Se_3$[16-20], iron-based superconductors[12-14,21-24], kagome superconductors[25,26], quasi-one-dimensional $K_2Cr_3As_3$[27], heavy fermion systems such as $UPt_3$[28], $UTe_2$[29], and, more recently, noble-metal-based superconductor $AuSn_4$[30,31]. Despite these advances, the realization of intrinsic topological superconductivity in bulk crystalline materials still remains a major challenge due to the scarcity of suitable material platforms and lack of robust experimental fingerprints beyond zero-energy excitations.

In recent years, crystalline-symmetry-protected route toward topological superconducting phases has attracted increasing attention as a promising strategy to address this challenge[32-36]. In particular, superconductors with nonsymmorphic crystal structures are especially compelling because their crystal symmetries, which combine point-group operations with fractional lattice translations, can enforce band degeneracies and give rise to nontrivial electronic structures such as hourglass dispersions and Dirac nodal lines[37-44]. In the presence of strong spin-orbit coupling, such symmetries can naturally generate locking between spin and other degrees of freedom on Fermi surfaces, imposing stringent constraints on the pairing tendency[45]. Nonsymmorphic superconductors have emerged as an especially appealing materials class for exploring exotic spin-triplet pairing states[45] and possible topological superconducting phases[45-52]. However, experimentally accessible nonsymmorphic superconductors with clear signatures of topological superconductivity remain rare.

Here, we report spontaneous rotational symmetry breaking and a robust zero-energy vortex bound state in the nonsymmorphic superconductor $PtPb_4$, which hosts a frustrated Shastry-Sutherland (SS) lattice and undergoes a superconducting transition at $T_c$=2.85 K. By employing both four-terminal and Corbino-like electrode configurations, both in-plane and out-of-plane *c*-axis resistivity measurements reveal pronounced twofold symmetry ($C_2$) of the superconducting state and upper critical field. Scanning tunneling microscopy/spectroscopy (STM/STS) characterizations further reveal elongated vortices aligned along specific crystallographic directions, providing direct real-space confirmation of the spontaneous rotational symmetry breaking of the underlying lattice. Remarkably, we observe a robust zero-energy vortex bound state, which persists without spatial splitting over extended distances, resembling the Majorana bound state.

## Results

The $PtPb_4$ crystallizes in tetragonal space group P4/nbm (No. 125), with lattice parameters $a$=$b$=6.73 Å, $c$=6.08 Å. Its crystal structure comprises alternating Pt and Pb layers, wherein Pt atoms are arranged into a square net and Pb atoms form a geometrically frustrated SS lattice within *ab* plane (Figure 1a). This unique structure exhibits nonsymmorphic symmetry, characterized by glide-mirror operations with fractional lattice translations, which gives rise to symmetry-protected nontrivial electronic structures such as Dirac nodal line crossings[39]. Consistently, the calculated topological invariants yield $Z_2 = 1$ at $k_z = 0$ and $Z_2 = 0$ at $k_z = \pi$, respectively, further confirming the strong topological character of $PtPb_4$ (Figure S1). High-quality $PtPb_4$ single crystals are synthesized by self-flux method. A typical optical photograph of a $PtPb_4$ single crystal reveals lateral dimensions of the millimeter scale (Figure 1b). Elemental analysis conducted on multiple samples exhibits an atomic ratio of Pt:Pb very close to 1:4 (Figure S2). The X-ray diffraction (XRD) pattern exclusively shows the (*00l*) reflections, indicating a pure phase oriented crystal surface (Figure S2). The double-crystal rocking curve of the (*001*) Bragg peak reveals a narrow full width at half maximum (FWHM) of 0.18° (Figure 1b). Moreover, the (*hk0*) diffraction pattern shows sharp, unsplit spots (Figure 1c), demonstrating high crystallinity and a low density of lattice defects. The refined lattice parameters $a$, $b$, and $c$ are 6.695 Å, 6.695 Å and 6.009 Å, respectively, which are in close agreement with theoretical values (Figure S3).

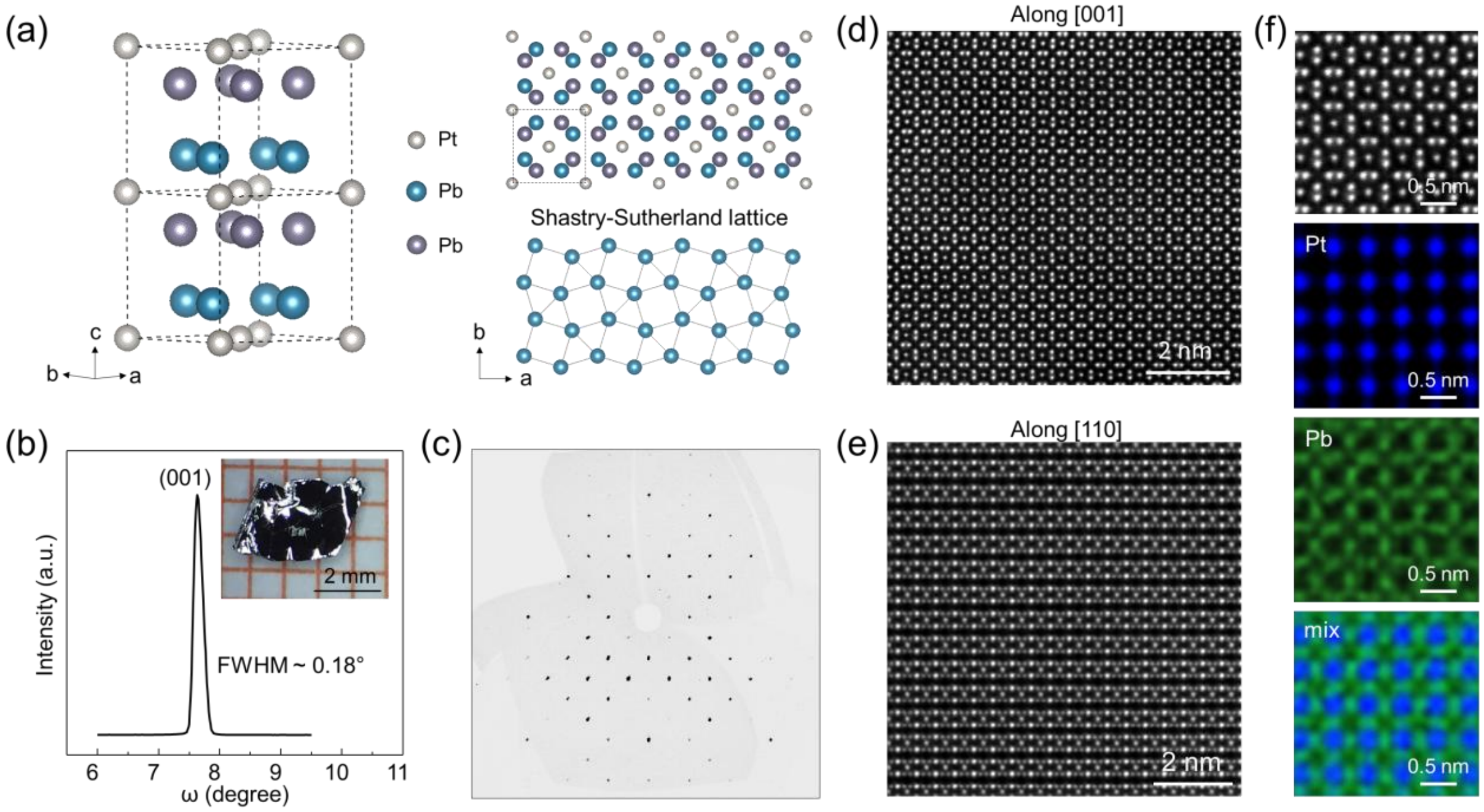

**Figure 1. Tetragonal structure of the nonsymmorphic $PtPb_4$.** (a) Crystal structure of $PtPb_4$, showing a tetragonal structure with $C_4$ symmetry. Two layers of Pb are sandwiched between consecutive layers of Pt. Each Pb layer forms a Shastry-Sutherland lattice while a square net is formed by each Pt layer, exhibiting a nonsymmorphic symmetry. (b) X-ray rocking curve of the (*001*) reflection, showing a narrow FWHM of 0.18°, indicative of high crystallinity. Inset shows an optical photograph of a $PtPb_4$ single crystal. (c) Diffraction pattern of the (*hk0*) plane, showing sharp diffraction spots without obvious splitting. (d,e) Atomic-resolution STEM images along the [001] and [110] directions, respectively, showing good agreement with simulated projections of the tetragonal structure. (f) Atomic-resolution EELS elemental mapping of Pt (blue) and Pb (green), acquired simultaneously with the ADF image. The false-color RGB map confirms distinct spatial distributions of Pt and Pb atoms.

We note an ongoing debate[53,54] regarding whether $PtPb_4$ is isostructural with $PtSn_4$, which exhibits AB-stacking with a slight structural distortion and crystallizes in an orthorhombic lattice with $C_2$ symmetry (Figure S4). To directly resolve the crystal structure of $PtPb_4$, we further perform atomic-scale structural and chemical analysis using aberration corrected scanning transmission electron microscopy (STEM). High-angle annular dark-field (HAADF) STEM images acquired along the [001] and [110] crystallographic directions (Figure 1d,e), unambiguously reveal a pristine tetragonal lattice, in excellent agreement with the structural models and the corresponding simulated STEM images (Figure S5). Moreover, atomic-resolution energy dispersive X-ray spectroscopy (EDS) mapping confirms the distinct spatial distributions of Pt and Pb atoms (Figure 1f), further verifying the chemical order. The absence of discernible structural defects, in complement with the XRD results, confirms the high crystallinity of the $PtPb_4$ crystal. The combined XRD and STEM analysis concludes that the $PtPb_4$ crystal crystallizes in a $C_4$ symmetry, distinct from the lower-symmetry orthorhombic phase of $PtSn_4$.

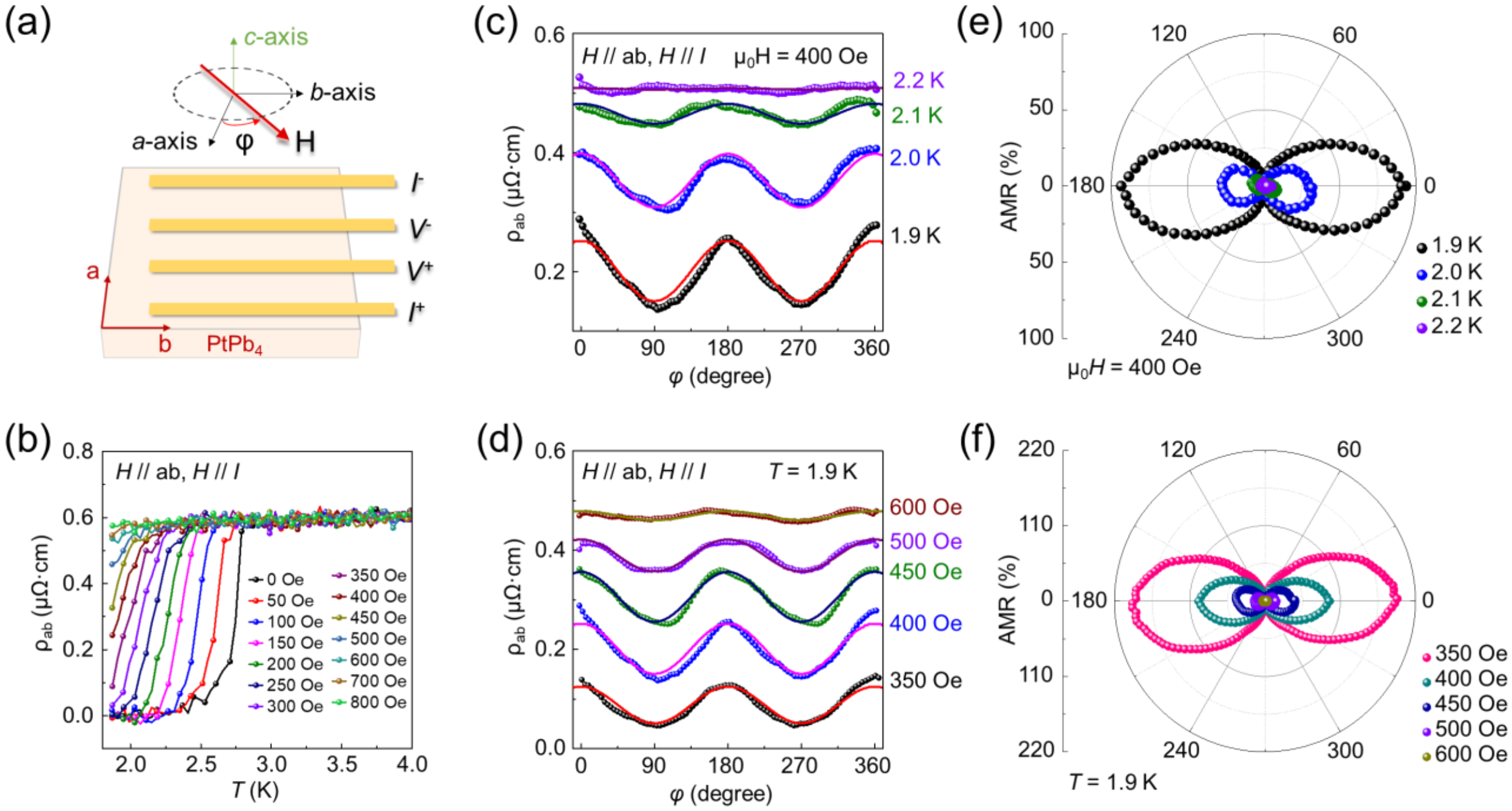

**Figure 2. In-plane twofold symmetric superconductivity of $PtPb_4$.** (a) Schematic of standard four-terminal electrode configuration for the in-plane resistivity measurements, with current along the *a*-axis and magnetic field rotated in *ab* plane, where φ denotes the angle between the in-plane magnetic field and crystallographic *a*-axis of $PtPb_4$. (b) Low-temperature $\rho_{ab}$ under in-plane magnetic fields ranging from 0 Oe to 800 Oe, showing a superconducting transition at $T_c = 2.85$ K. (c) Angle dependence of the in-plane resistivity $\rho_{ab}$ under 400 Oe at temperatures between 1.9 K and 2.3 K, showing that the twofold anisotropy weakens with increasing temperature. (d) Angular dependence of $\rho_{ab}$ at 1.9 K under magnetic fields from 350 Oe to 600 Oe, showing the twofold anisotropy weakens with increasing magnetic field. (e) Angular-dependent magnetoresistance plotted in polar coordinate measured at 1.9 K under in-plane magnetic fields of 350 Oe, 400 Oe and 450 Oe, showing a pronounced anisotropy with an AMR ratio of 194% at 350 Oe. (f) Angular-dependent magnetoresistance plotted in polar coordinate measured at 1.9 K to 2.2 K under a fixed magnetic field of 400 Oe, showing that the twofold anisotropy weakens with increasing temperature.

We then study the superconductivity in the tetragonal $PtPb_4$ by performing angle-dependent resistivity measurements using a standard four-probe configuration (Figure 2a). The in-plane resistivity $\rho_{ab}$ exhibits a sharp superconducting transition that occurs at the onset temperature $T_c \sim 2.85$ K (Figure S6). Upon applying in-plane magnetic fields, the transition shifts gradually to lower temperatures and the superconductivity is almost suppressed up to 600 Oe at 1.9 K (Figure 2b). Strikingly, the azimuthal-angle $\varphi$-dependent in-plane resistivity in superconducting regime exhibits pronounced $C_2$ oscillations under rotating magnetic field, following a cos2$\varphi$ dependence (Figure 2c,d), suggesting an in-plane twofold symmetry of the superconductivity in $PtPb_4$. This twofold symmetry is well presented in polar plots (Figure 2e,f), which displays a dumbbell-like shape with a maximum anisotropic magnetoresistance (AMR) of ~194%. With increasing magnetic field or temperature, this anisotropy gradually weakens and eventually vanishes in the normal state (Figure 2c-2f). These results suggest an unconventional pairing state in $PtPb_4$ that would break the underlying $C_4$ lattice symmetry.

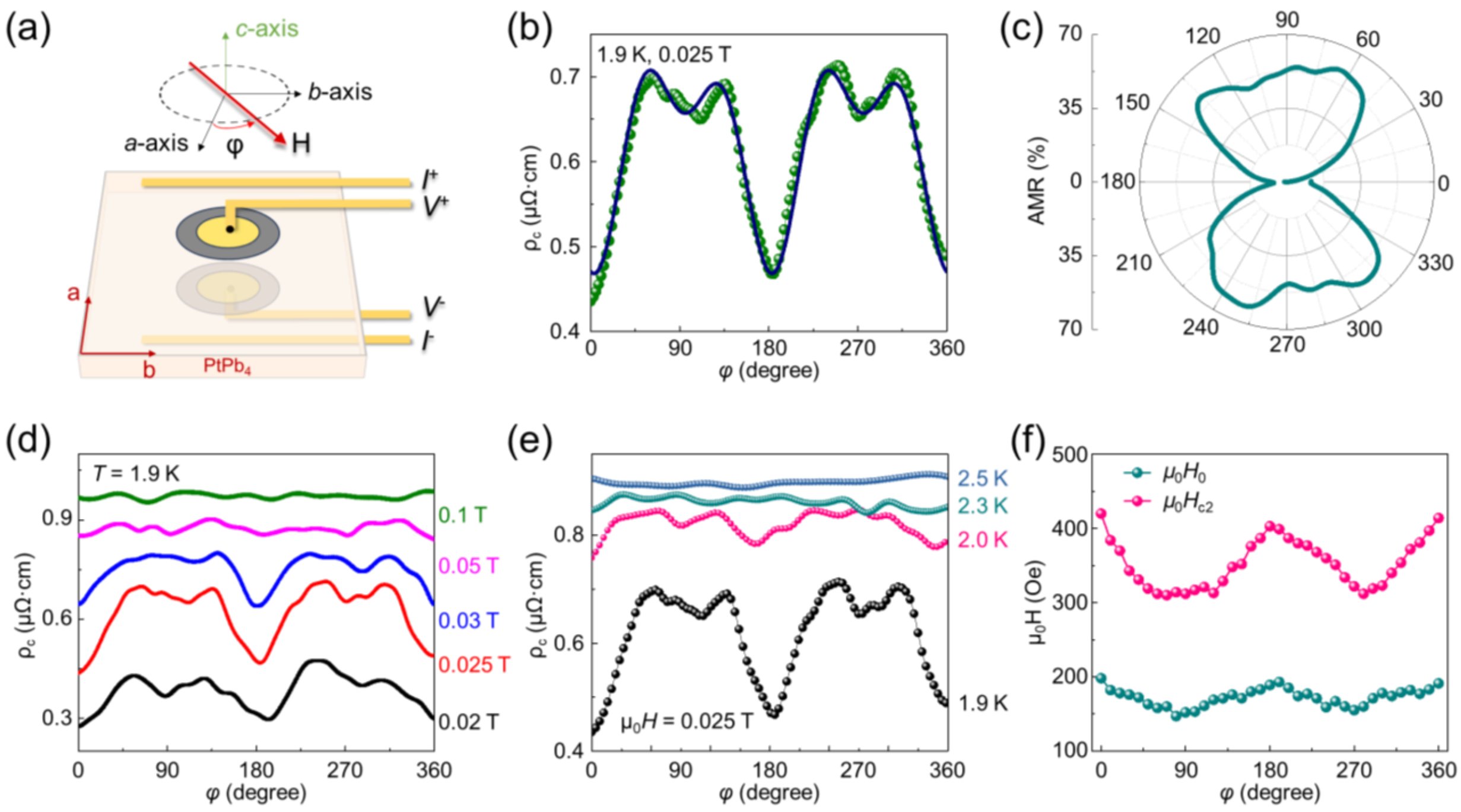

**Figure 3. Twofold symmetric superconductivity in $PtPb_4$ from out-of-plane resistivity.** (a) Schematic of a Corbino-shape-like electrode configuration for the *c*-axis resistivity measurements, where current flows along the *c*-axis and magnetic field is initially aligned with *a*-axis and rotated within the *ab* plane. (b) Angular dependence of the *c*-axis resistivity at 1.9 K under 0.025 T, fitted with a trigonometric function, showing a $C_2$ symmetry, indicating rotational symmetry breaking of the superconductivity. (c) Polar plot of the angular dependent *c*-axis resistivity at 1.9 K under 0.025 T, showing pronounced twofold anisotropy with an AMR ratio of 70%. (d) Angular dependence of $\rho_c(\varphi)$ at different in-plane magnetic fields, showing the twofold symmetry weakens with increasing magnetic field. (e) Angular dependent *c*-axis resistivity measured at different temperatures under 0.025 T, showing the twofold symmetry weakens with increasing temperature. (f) Angular dependent upper critical field ($\mu_0 H_{c2}$) and zero-resistance field ($\mu_0 H_0$), both showing twofold symmetry.

In order to evidence the intrinsic $C_2$-symmetric superconductivity in $PtPb_4$, we design a Corbino-like contact geometry, in which the current flows predominantly along the *c*-axis and remains perpendicular to the rotating in-plane magnetic field (Figure 3a). Temperature-dependent *c*-axis resistivity ($\rho_c$) shows a large anisotropy compared to in-plane resistivity $\rho_{ab}$ (Figure S7). A clear superconducting transition at 2.85 K is also observed and it is gradually suppressed with increasing magnetic field up to 600 Oe at 1.9 K, consistent with the in-plane resistivity measurements. Remarkably, the angular dependence of $\rho_c$ in the superconducting regime (1.9 K, 0.025 T) exhibits obvious $C_2$ symmetry with the local minima near $\varphi = 0°$ and 180°, corresponding to the *a*-axis direction (Figure 3b,c). The maximum AMR is about 70%, significantly smaller than that measured from in-plane resistivity. The minimum resistivity reflects a relatively larger upper critical field $\mu_0 H_{c2}$ along the *a*-axis, indicating the existence of intrinsic $C_2$ symmetry of superconductivity that breaks the rotational symmetry in $PtPb_4$.

Notably, except $C_2$ anisotropy, extra oscillation can also be seen in the $\rho_c$ curve. The Fourier transformation (FT) of the $\rho_c$ curve (Figure S8a) exhibits a strong FT amplitude ($A_{FT}$) peak at 180° and a small peak at 90°, indicating the coexistence of $C_2$ and $C_4$ anisotropy, which is possibly originating from superconductivity and vortex dynamics. At low fields near the transition (e.g., 0.025 T), vortex flow produces twofold anisotropy in *c*-axis resistivity, consistent with a $C_2$-symmetric vortex structure. In a tetragonal lattice, such vortices can align along both the *a*- and *b*-axes, forming domains that give rise to an additional $C_4$ component.

In addition, with increasing the magnetic field or temperature, such twofold symmetry is progressively suppressed (Figure 3d, 3e, S8b), further indicating that the symmetry breaking arise from the superconducting state. Moreover, both the angle-dependent upper critical field $\mu_0 H_{c2}$ and near zero-resistance field $\mu_0 H_0$ exhibit a pronounced $C_2$ symmetry with maxima at $\varphi = 0°$ and 180° (Figure 3f), coinciding with the resistivity minima. The in-plane anisotropy ratio of $\mu_0 H_{c2}$ is approximately 1.4, consistent with that extracted from temperature-dependent resistivity measurements. These results provide compelling evidence for intrinsic rotational symmetry breaking of the superconducting state in the topological $PtPb_4$.

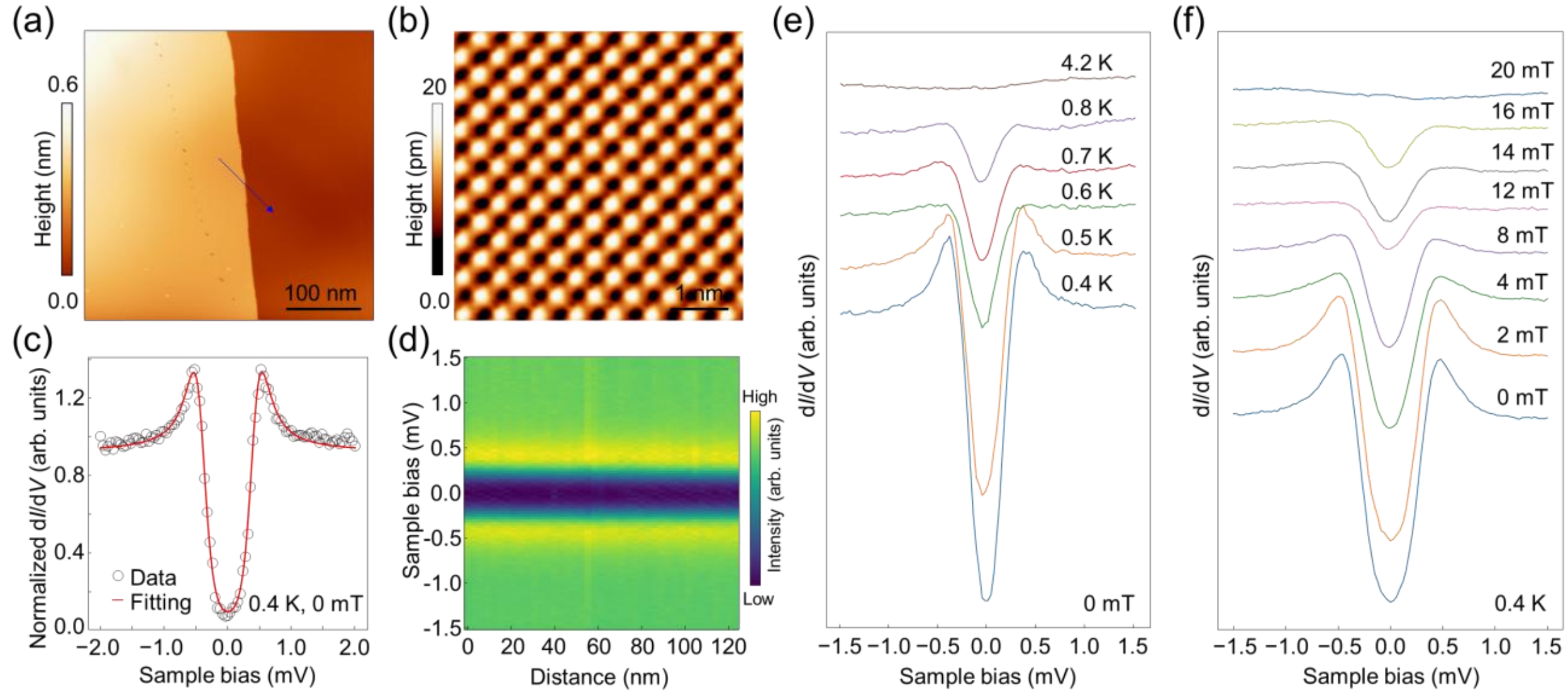

**Figure 4. Temperature and magnetic-field evolution of superconductivity on $PtPb_4$ surface.** (a) Large-scale STM topography (400 nm × 400 nm) of the $PtPb_4$ surface ($V_s$=-500 mV, $I_t$=20 pA), showing atomically flat terrace with a step height of ~600 pm. (b) Atomically-resolved STM image acquired at superconducting state (0.4 K), showing square atomic lattice preserving $C_4$ symmetry. (c) A representative d$I$/d$V$ spectrum acquired on $PtPb_4$ surface, showing a superconducting gap Δ~0.42 meV and symmetric coherence peaks. (d) Intensity map of a d$I$/d$V$ linecut along the blue arrow in (a), demonstrating global homogeneous superconductivity in $PtPb_4$. (e,f) Temperature and out-of-plane magnetic-field dependent d$I$/d$V$ spectra (vertically offset for clarity), showing the evolution of superconductivity with temperature and magnetic field, with estimated critical temperature of ~2.8 K and critical magnetic field of ~20 mT. Setpoint: $V_s$=-1.5 mV, $I_t$=400 pA, $V_{mod}$=50 μV.

We further investigate the symmetry-broken superconducting state by direct imaging of quasiparticle excitations in the vicinity of magnetic vortex cores using ultralow temperature STM/STS. Large-scale STM topography of a freshly cleaved $PtPb_4$ crystal shows an atomically flat terrace (Figure 4a). Atomically resolved imaging at superconducting state reveals a perfect Pb-terminated square lattice that retains $C_4$ rotation symmetry (Figure 4b, S9), consistent with the STEM results, indicating no structural transition in $PtPb_4$. A representative normalized differential conductance (d$I$/d$V$) spectrum exhibits a superconducting gap with Δ~0.42 meV fitted by Dynes model (Figure 4c). A d$I$/d$V$ linecut across the terrace further reveals uniform spectra without any detectable contrast at the step edge, evidencing the robustness of superconducting gap (Figure 4d). Temperature-dependent d$I$/d$V$ spectra trace the gradual suppression of the superconducting gap when the temperature increases to $T_c$~2.8 K (Figure 4e). In addition, a critical out-of-plane $\mu_0 H_0$~20 mT is estimated from the field evolution of the d$I$/d$V$ spectra (Figure 4f), consistent with the transport results (Figure 2b).

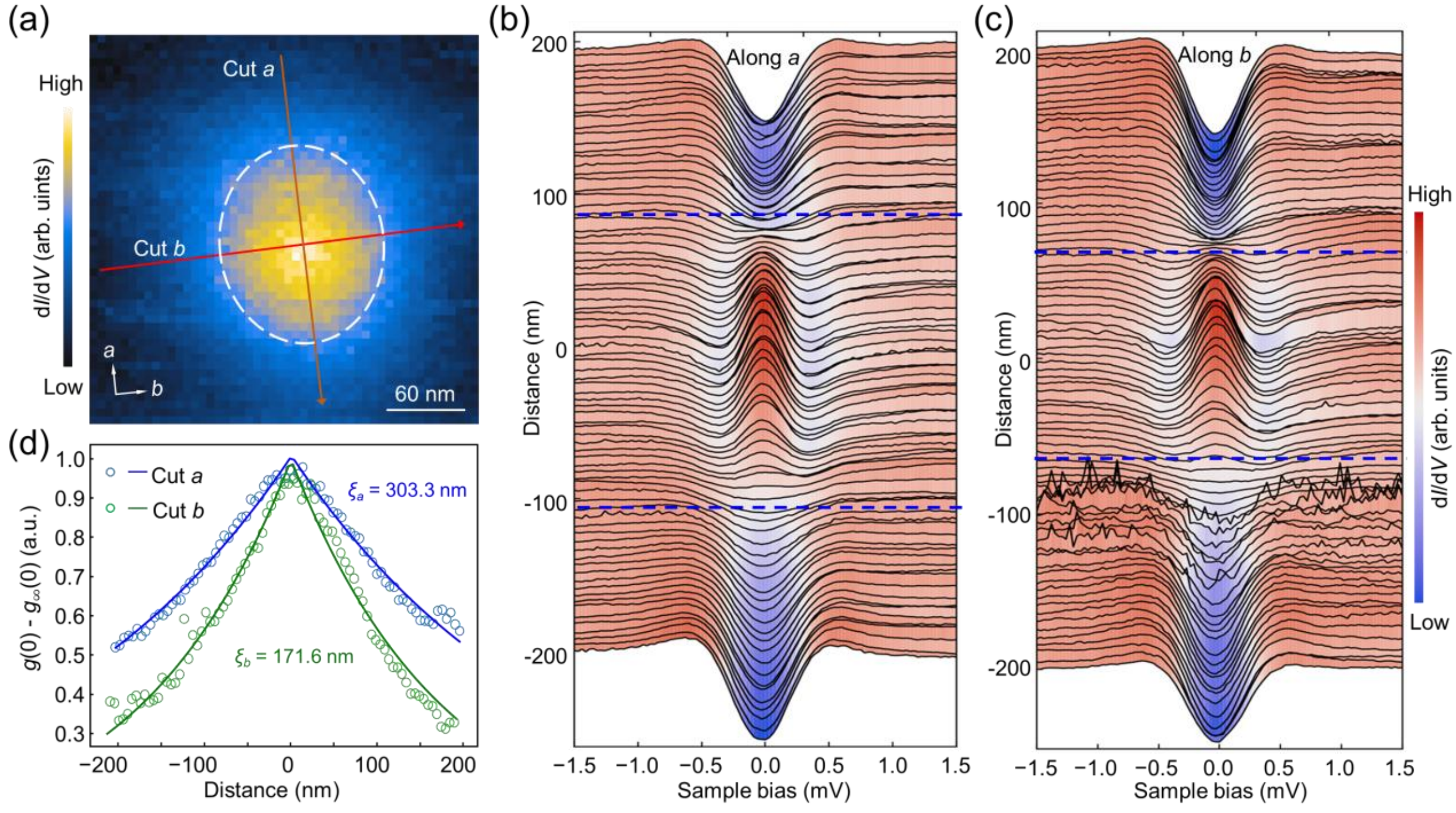

**Figure 5. Rotational symmetry breaking of vortex core states in $PtPb_4$.** (a) Zero-energy spectroscopic imaging of an elliptical vortex, outlined by white dashed ellipse, showing an anisotropic superconducting vortex core elongated along the *a*-axis. (b,c) The d*I*/d*V* linecuts along Cut *a* and Cut *b*, respectively, approximately aligned with the major (b) and minor (c) axes of the elliptical vortex, showing anisotropic vortex bound states. Blue dashed-lines serve as guides to the eye for the qualitative view of the anisotropic vortex bound states. (d) Exponential fitting (solid curves) of zero-energy conductance linecuts along Cut *a* (blue circles) and Cut *b* (green circles), showing the coherence length $\xi_a$ = 303.3 nm, $\xi_b$ = 171.6 nm with anisotropic ratio of the vortex $\xi_a/\xi_b$~1.8 core. Setpoint: $V_s$=-1.5 mV, $I_t$=400 pA, $V_{mod}$=50 μV.

To directly visualize vortex state, we performed spectroscopic imaging under a perpendicular field of ~10 mT, half the critical value. Despite the $C_4$-symmetric atomic lattice, intriguingly, the zero-bias conductance map reveals an elliptical vortex (Figure 5a). The long axis of the elliptical vortex aligns with *a*-axis of $PtPb_4$, whereas the short axis aligns with *b*-axis, thereby reducing the rotational symmetry from $C_4$ to $C_2$. False-color linecut intensity maps taken along *a*-axis and *b*-axis accentuate the pronounced elongation of the vortex core (Figure 5b,c), which exhibits a robust zero-energy bound state persisting without spatial splitting over extended distances, closely resembling a Majorana bound state. Blue dashed-lines serve as guides to the eye for the qualitative view of the anisotropic vortex bound states. Quantitative analysis is achieved by fitting the zero-bias conductance profiles along Cut $a$ and Cut $b$ to the exponential form $g_0(r) = g_0(\infty) + A\mathrm{e}^{-r/\xi}$ (Figure 5d). The results yield coherence lengths of $\xi_a$ = 303.3 nm along $a$-axis and $\xi_b$ = 171.6 nm along $b$-axis, giving a pronounced anisotropy ratio $\xi_a/\xi_b$~1.8 that directly reflects the rotational symmetry breaking of the superconducting state in $PtPb_4$.

Finally, we confirm the alignment of the elongated direction of superconducting vortices with a crystallographic direction. Figures S10 displays two representative elliptical vortices whose major axes are alternately aligned with $b$- (Figure S10a) and $a$-axes (Figure S10d), respectively. This mutually orthogonal orientation of the elliptical cores, observed on the terraces of the same crystal, confirms that the fourfold rotational symmetry of the underlying lattice is spontaneously broken within the superconducting state. Quantitative exponential fits to the zero-bias conductance profiles again yield anisotropic coherence lengths, corroborating the pronounced elongation of the vortex cores along the selected crystallographic axes.

## Discussions

Through combined measurements of resistivity, upper critical field, and vortex morphology, we have revealed intrinsic rotational symmetry breaking in the superconducting state of the nonsymmorphic superconductor $PtPb_4$. The microscopic origin of this symmetry breaking, however, remains an intriguing open question. As neither obvious structural distortion nor density wave order is observed, one natural possibility is the emergence of nematic superconductivity[17-19,55,56]. The superconducting state of $PtPb_4$ may have a multi-component order parameter belonging to the two-dimensional representation whose degeneracy is spontaneously lifted or two nearly degenerate order parameters belonging to different one-dimensional representations which compete and mix with each other, lowering the rotational symmetry of the condensate while leaving the crystal lattice itself unchanged. Within this scenario, the unique crystal structure of the system may be an important ingredient. The nonsymmorphic crystal of $PtPb_4$ with local-inversion symmetry breaking[57], in the presence of strong spin-orbit coupling, can lead to topological nontrivial electronic structure such as Dirac nodal line crossings[39] and may stabilize multi-component odd-parity superconducting state[44,58,59]. Such pairing state, analogous to the case in $Cu_xBi_2Se_3$, may delicately generate a topological superconducting phase supporting Majorana zero modes in vortices[18]. Therefore, the combination of spontaneous rotational symmetry breaking in superconductivity and the presence of robust zero-energy vortex bound state suggests that $PtPb_4$ may host an unconventional superconducting phase intertwined with nontrivial topology, providing a fertile platform for exploring exotic pairing mechanisms and emergent Majorana physics.

## Conclusions

In summary, we identify $PtPb_4$ as a nonsymmorphic superconducting platform hosting spontaneous rotational symmetry breaking and a robust zero-energy vortex bound state. The pronounced twofold anisotropy in transport and the real-space visualization of elongated vortices provide compelling evidence for a symmetry-broken superconducting state. The unsplit zero-energy vortex bound state further supports nontrivial topological superconductivity and possible Majorana physics. More broadly, the coexistence of nonsymmorphic symmetry, frustrated lattice geometry, and nontrivial electronic topology, $PtPb_4$ offers a unique setting for exploring unconventional superconductivity and developing future topological superconducting and quantum-device architectures.

## Methods

**Materials synthesis.** High-quality $PtPb_4$ single crystals were synthesized using a self-flux method. Platinum and lead in a molar ratio of 1:7 were thoroughly mixed and sealed in an evacuated quartz ampoule with a flame torch. The sealed tube was heated to 700 °C over 12 hours, held for 48 hours, and then slowly cooled to 320 °C at a rate of 1 °C per hour. At 320 °C, the excess flux was removed by centrifugation, yielding high-quality, millimeter-scale single crystals.

**Structural characterizations.** XRD patterns were taken by a Rigaku SmartLab SE X-ray diffractometer with Cu Kα radiation (λ= 0.15418 nm) at room temperature. A Bruker D8 VentureSingle was adopted for obtaining crystal diffraction patterns. Scanning electron microscopy (SEM) and X-ray energy-dispersive spectroscopy (EDS) were performed using a HITACHI S5000 with an energy dispersive analysis system Bruker XFlash 6|60. The cross-sectional sample for STEM analysis was prepared using a Thermofisher Helios G4 CX focused ion beam system (FIB). Atomic-scale STEM imaging and EDS spectrum imaging were carried out on an aberration-corrected JEOL GrandARM2 microscope with an energy dispersive analysis system JEOL JED-2300T, operated at an acceleration voltage of 200 kV.

**Resistivity measurements.** Resistivity measurements were performed in a physical property measurement system (PPMS, Quantum Design). Single-crystal diffraction was used to determine the crystallographic orientations, along which the samples were cleaved and edges cut. The in-plane resistivity was measured using a standard four-terminal configuration with current flowing within the *ab*-plane, whereas the *c*-axis resistivity was measured using a Corbino-like geometry. To eliminate the influence of residual Hall components on the angular-dependent resistivity data, measurements at each angle were averaged over positive and negative magnetic fields.

**Scanning tunneling microscopy/spectroscopy.** The $PtPb_4$ crystal was introduced in a UHV chamber, then cleaved by aluminum rod, and immediately transferred to a low-temperature STM head. The base pressure during this procedure was $1\times10^{-10}$ mbar. STM measurements were performed in a commercial STM working at 0.4 K. Tungsten tip was fabricated via electrochemical etching and calibrated on a clean Au(111) surface prepared by repeated cycles of sputtering with argon ions and annealing at 770 K. All STM images were taken in constant-current mode and the differential conductance (d$I$/d$V$) spectra were taken using a lock-in amplifier (f = 973.123Hz).

**DFT calculations.** First-principles calculations based on density functional theory (DFT) were carried out with the Vienna ab initio simulation package (VASP)[60]. The projector augmented wave (PAW) method is used to describe the interaction between core and valence electrons[61]. The electron exchange-correlation functional is dealt with the generalized gradient approximation (GGA) in the PBE form[62]. The orbital-dependent on-site Coulomb interactions (U) is considered with the values of 3 eV and the cutoff energy of plane-wave is taken as 500 eV, and the total energy convergence threshold is $10^{-6}$ eV/atom. The full structure optimizations on atomic positions and lattice vectors are performed until the maximum force on each atom is less than 0.001 eV/Å. The 5 × 5 × 5 and 9 × 9 × 9 K-meshes generated by a Γ-centered Monkhorst-Pack grid are used for structure optimization and self-consistent calculations. Topological properties are investigated by an effective tight-binding Hamiltonian constructed from the maximally localized Wannier functions. The iterative Green function method is used with the package WANNIERTOOLS[63].

**Author Contributions.** H.C., H.G. and H.-J.G. designed the experiments. H.G. and S.L. prepared the samples. S.L., H.G., K.Z., and H.Y. performed the transport measurements. Z.L., Z.H., S.L., X.H., X.L., and H.C. performed the STM/S measurements. T.G. and W.Z. performed the STEM measurements. K.H. and L.Z. carried out the theoretical calculations. All authors participated in the data analysis and manuscript writing.

**Acknowledgements.** The authors are grateful to Ziqiang Wang, Hengxin Tan, and Haijing Zhang for valuable discussions. The work is supported by grants from the National Key Research and Development Projects of China (2022YFA1204100), the National Natural Science Foundation of China (62488201, 52572188, 92580202), the Chinese Academy of Sciences (YSBR-053, YSBR-003).

**Data availability.** The data that support the findings of this article are openly available from the corresponding authors on request.

**Competing interests.** The authors declare that they have no competing interests.